\documentstyle[aps,preprint,epsf]{revtex}
\tighten

\newcommand{\beq}{\begin{equation}}
\newcommand{\beqn}{\begin{eqnarray}}
\newcommand{\eeq}{\end{equation}}
\newcommand{\eeqn}{\end{eqnarray}}
\newcommand{\vp}{\varphi}
\newcommand{\dvp}{\delta\varphi}
\newcommand{\ts}{\textstyle}
\newcommand{\rd}{\displaystyle{\cdot}}
\begin{document}
\draft

\preprint{HUTP-98/A058, PU-RCG-98/11; hep-ph/9808404}

\title{General relativistic effects in preheating}

\author{Bruce A. Bassett$^{\dagger}$, David I.
Kaiser$^{*}$, and
Roy  Maartens$^{\ddagger}$}

\address{~}
\address{$\dagger$ Department of Theoretical Physics,
Oxford University, Oxford~OX1~3NP, England}
\address{$^*$  Lyman Laboratory of Physics, Harvard University,
Cambridge, MA~02138 USA}
\address{$\ddagger$ School of Computer Science
and Mathematics, Portsmouth University,
Portsmouth~PO1~2EG, England}
\address{~}
\date{\today}
\maketitle

\begin{abstract}

General relativistic effects in the form of metric perturbations are
usually neglected in the preheating era that follows inflation.  We argue
that in realistic multi-field models these effects are in fact crucial,
and the fully coupled system of metric and quantum field fluctuations
needs to be considered. Metric perturbations are resonantly amplified,
breaking the scale-invariance of the primordial spectrum, and in turn
stimulate scalar field resonances via gravitational rescattering. This
non-gravitationally dominated nonlinear growth of gravitational
fluctuations may have significant effects on the Doppler peaks in the
cosmic background radiation, primordial black hole formation,
gravitational waves and nonthermal symmetry restoration.

\end{abstract}
\vskip 2pc
\pacs{98.80.Cq \hspace*{2cm}
hep-ph/9808404}

As recent ideas about the end of an inflationary era have shown,
post-inflation reheating was one of the most violent and
explosive processes occurring in the early universe
\cite{preh,exp,KLS}. The
nonequilibrium, nonperturbative, resonant decay of the inflaton
was demonstrated in Minkowski
spacetime.  The first steps toward a
gravitationally self-consistent treatment,
which modeled inflaton
decay in an expanding, dynamical background spacetime,
revealed that
``preheating" may proceed with qualitative, as well as quantitative,
differences from the Minkowski case \cite{exp,KLS}. Yet even these
studies neglected an essential feature of gravitational physics:
the
production and amplification of metric perturbations attending such a
sudden transfer of energy from the oscillating inflaton to
higher-momentum particles.
Other papers began the study of metric perturbations
induced during preheating: in \cite{metric}
a perfect-fluid analysis with Born decay was used, and crucial
features of preheating were thus not incorporated;
in \cite{NT} these limitations are avoided, but
the effects of the amplified metric perturbations on the process of
preheating itself are not considered.

In this Letter, we pursue a more self-consistent
relativistic treatment of
preheating, by studying both the field fluctuations (responsible for
particle production) and the coupled metric perturbations (describing
gravitational fluctuations in the curvature).
In the process, we clarify the question
of causality and the amplification of long-wavelength perturbations
during preheating.
We give qualitative arguments, confirmed by numerical
results (see also \cite{BKMT}),
to show that {\em metric perturbations typically undergo
rapid growth during multi-field
preheating, and in turn act as a source and a pump for the growth of
field fluctuations via gravitational rescattering,}
ultimately making the preheating process
even more efficient than previously realized.
Neglecting this general relativistic effect can produce
misleading results.

Moreover,
nonlinear growth of metric perturbations is itself
of potentially major significance, since it precipitates
nonlinear density fluctuations, leading to strong mode-mode
coupling effects, and nonlinear deviations
from the conformally flat background.
This could affect observable quantities such as the
cosmic microwave background (CMB)
spectrum, complicating the usual predictions from
inflationary models. Gravitational wave power could be
enhanced by gravitational bremstrahlung, and primordial
black hole production could occur without the need for
special properties in the power spectrum.

We work with the gauge-invariant formalism of \cite{MFB} to study
the evolution of scalar perturbations of the metric. (The
amplification of gravitational waves at preheating has been
considered in \cite{gw}.) For scalar perturbations, in the case of
a scalar-field energy-momentum tensor and spatially flat
background, the perturbed metric in the
longitudinal gauge is $ds^2 = a^2 (\eta) \left[ \left( 1 + 2 \Phi
\right) d \eta^2 - \left( 1 - 2\Phi \right)d{\bf x}^2
\right]$,
where $\Phi$ is the gauge-invariant gravitational potential.

Realistic reheating models involve at least one field coupled
to the inflaton $\vp(\eta)$. First we consider models with only an
inflaton, in order to give a simple illustration of some effects.
The gauge-invariant field fluctuations
$\delta\varphi$ and the metric perturbations
$\Phi$ obey the coupled Eqs. (6.40-42) of
\cite{MFB}.  Defining the rescaled
fields $Y \equiv a\Phi$ and $X \equiv a
\delta \varphi$, these become,
in momentum space:
\beqn
Y_k^{\prime\prime} + \left[ k^2 - {\textstyle{1\over2}}{\kappa^2}
\varphi^{\prime \> 2} \right] Y_k &=& \kappa^2
\varphi^{\prime\prime} X_k \,,
\label{Yeq1}\\
X_k^{\prime\prime} + \left[ k^2 + a^2 V_{\varphi\varphi} -
{\textstyle{3\over2}}
\kappa^2 \varphi^{\prime \> 2} - 2 {\cal H}^2 \right] X_k
&=& 2 \varphi^{\prime\prime} Y_k \,,
\label{Xeq1}\\
Y_k^\prime &=& {\textstyle{1\over2}}\kappa^2
\varphi^\prime X_k \,, \label{y0i}
\eeqn
where $V(\vp)$ is the potential,
${\cal H}= a'/a$ and $\kappa^2= 8 \pi M_{\rm pl}^{-2}$.

Eq. (\ref{y0i}) is a constraint, showing
that if there is
explosive growth in the field fluctuations
$X_k$ -- the  heart of preheating --
then this automatically constrains the gravitational
fluctuations $Y_k$ to
follow, and {\em vice versa}.
This equation essentially says that if you ``shake" the
right-hand side of Einstein's field equations
$G_{\mu\nu}=\kappa^2 T_{\mu\nu}$, then unavoidably
you are simultaneously ``shaking" the left hand side.
Clearly, neglecting the metric perturbations $Y_k$ can
be seriously misleading under many conditions, since it
is tantamount to ignoring the perturbed
Einstein equations.
This may be reasonable in a slow-roll inflationary
regime, but generally not in an oscillatory regime.

A clear illustration is provided by super-Hubble modes in the simplest
model, $V={1\over2}m^2\vp^2$. If $\dvp_*$ is the field fluctuation
calculated by neglecting metric perturbations, then for $k\rightarrow 0$,
$X_*\equiv a\dvp_*$ satisfies Eq. (\ref{Xeq1})  with gravitational
fluctuations eliminated, i.e.
$X_*''+\left[m^2a^2+{1\over2}\kappa^2\vp'^2-2{\cal H}^2\right]X_*=0$.
This equation is then of precisely the same form as the {\em background}
Klein-Gordon equation, so that
\[
\dvp_*\propto\vp\approx\vp_0\sin(b\eta^3)/ (b\eta^3)\,,
\]
where $\vp_0$ and $b$ are constants. The
approximation arises from using the time-averaged scale factor
$\bar{a}\propto\eta^2$, and improves in accuracy as $\eta$ increases.
When metric perturbations are incorporated, the long-wavelength solution
is given in general by [see \cite{MFB}, Eq. (6.57)]
$\dvp\propto\vp'a^{-2}\int a^2d\eta$.  We find that at reheating
\[
\dvp\approx {\textstyle{3\over5}}\vp_0\cos(b\eta^3)\,,
\]
to lowest order in
$\eta^{-1}$.  Gravitational rescattering produces a non-decaying term in
the field fluctuations, which dominates the rapidly decaying fluctuation
$\dvp_*$ calculated by neglecting gravitational fluctuations.  Relating in
the usual way \cite{preh,KLS} the field fluctuations to the density of
particles produced per mode, we find that this non-decaying solution
indicates nonzero particle production.  If gravitational perturbations are
neglected, {\em no} particle production is found for this model. {\em
Gravitational rescattering dramatically alters the evolution of the
matter-field fluctuations even in this simplest of all models, especially
on super-Hubble-radius scales.} Since preheating concerns primarily the
behavior of such matter-field fluctuations after inflation, it is thus
crucial to study the coupled metric perturbation - field fluctuation
system.

Now we consider the amplification of gravitational fluctuations.
The constrained system of equations (\ref{Yeq1})--(\ref{y0i})
has one degree of freedom, reflected in the decoupled equation
$Y_k''-2\left(\vp''/\vp'\right)Y_k'-\left({\textstyle{1\over2}}
\kappa^2\vp'^2-k^2\right)Y_k=0 $.
To avoid the periodic singularities when $\varphi' = 0$,
Nambu and Taruya \cite{NT} employ the rescaled Mukhanov variable
$\tilde{Q}=a^{(2-n)/(n+1)}\left[X+(\vp'/{\cal H})Y\right]$, where
$n$ is the index in the power-law potential
$V={1\over2}m^2\vp_0^2(\vp/\vp_0)^{2n}$. Using the
time-averaged forms of $a$ and $\vp$,
they find that, to leading order in
$\tau^{-1}\propto\eta^{3/(1-2n)}\propto a^{-3/(n+1)}$,
\beq
\tilde{Q}_{\tau\tau}+\left[1+ k^2\gamma^2\tau^{4(n-2)/3}-(4/\tau)
\sin 2\tau\right]\tilde{Q}=0 \,,
\label{nt}\eeq
where $\gamma$ is a constant. This equation has Mathieu form,
$y''+[A_k-2q\sin 2x]y=0$, with time-dependent $A_k$ and $q$, so that modes
can be drawn through instability bands by the expansion of the universe,
if they were not already there initially \cite{NT,Kandrup}.
Note that
the resonance parameters scale as $q/A_k \propto
a^{-3/(n+1)}\left[\beta^2+k^2a^{4(n-2)/(n+1)}\right]^{-1}$,
where $\beta$ is constant.
If metric fluctuations are neglected, then
$q/A_k \propto a^{-3}$.
Therefore, for $n<8$, {\em expansion is less effective in
ending resonance
when metric fluctuations are incorporated.}

Further discussion of single-field models is given in \cite{FB,PE}, which
confirm the analytical conclusions of \cite{NT}:  in the $n = 1$
(quadratic potential) model, there is no resonance in the long-wavelength
limit, whereas large, resonant growth is found for other models, such as
the massless, quartically-coupled case ($n = 2$). However, as
pointed out above, the single-field case is completely inadequate
as a model of reheating, and we turn now to consider the
multi-field case.

The rapid growth of metric perturbations in the multi-field case (see
Fig. 1) will produce a backreaction on
the background quantities $a$ and $\varphi$ \cite{ABM}.
Similarly, the amplified field fluctuations will grow to
be of the same order as the tree-level terms, such as $\varphi^2$,
and hence will damp the inflaton's oscillations, ending the
parametric resonance.
Let us consider an effective single-field model.
We may then estimate these two time-scales, expecting the initial
preheating phase to end at $\eta_{\rm end} = {\rm min} \{ \eta_{\rm m},\>
\eta_{\rm f } \}$, where $\langle \Phi^2
(\eta_{\rm m}) \rangle = 1$ and $\langle \delta \varphi^2
(\eta_{\rm f})\rangle = \varphi^2$.
In a resonance band, $\Phi_k=F_k(\eta)e^{\mu_k\eta}$,
where $F_k$ oscillates and $\mu_k$ is the Floquet exponent. From Eq.
(\ref{y0i}), and working in the saddle-point approximation,
we may estimate
\[
\langle \Phi^2 \rangle=(2\pi)^{-3}
 \int d^3 k \vert \Phi_k \vert^2\sim \kappa^4 \varphi^{\prime
\> 2} \langle \delta \varphi^2 \rangle / (4 \mu_{k_{\rm max}}^2)\,,
\]
where $\mu_{k_{\rm max}}$ is the maximum Floquet exponent.
For single-inflaton models of chaotic inflation with polynomial
potentials, the slow-roll conditions are violated, and the inflaton
begins to oscillate at $\varphi_0 = \alpha M_{\rm pl}$, with
$\alpha \sim 0.3$.
In models with a massive inflaton, we
may further approximate $\vert \varphi^\prime \vert \sim \alpha
m M_{\rm pl}$.
The field fluctuations saturate their (linear-theory) upper
limit at $\langle \delta \varphi^2 \rangle \sim \varphi^2$.  Combining
these  yields $\langle \Phi^2 (\eta_{\rm m}) \rangle \sim 16 \pi^2
\alpha^4 m^2/\mu_{k_{\rm max}}^2$.  The specific
spectrum of fluctuations, governed
by the values of $\mu_k$, depends on details of the
potential.  For modes subject to a parametric resonance,
$\mu_{k_{\rm max}} \sim \alpha m$,
whereas modes subject to a
negative-coupling instability may have
$\mu_{k_{\rm max}} \sim O(m)$  \cite{preh,exp,KLS,negg}.  Thus
first-order analysis reveals that gravitational backreaction will
become
relevant at around the same time as the
backreaction of the nonlinearly-coupled field
fluctuations.

Another important issue is to demonstrate how super-Hubble modes
may be amplified causally. In inflation, modes which had been deep within
the Hubble radius {\em during} inflation become amplified and stretched to
super-Hubble scales \cite{MFB,inflreviews}.  Reheating was
believed not to be able to affect these super-Hubble scales.  In
preheating, however, the coherence of the inflaton condensate immediately
after inflation {\em does} allow for super-Hubble dynamics.  At first,
such behavior might appear to violate causality.  This is not the case;
consider the following:  (1) the field equations are relativistic, hence
causality is automatically built into the solutions; (2) causality
concerns space-like related events, and does not translate into direct
constraints on individual modes in Fourier space \cite{causal}; (3)
explicit calculation
of the unequal-time two-point correlation function reveals that no mass or
energy is being transported superluminally by these super-Hubble
resonances.  See \cite{BKMT} for this calculation and further discussion.
Similar conclusions regarding the possibility for the causal amplification
of super-Hubble modes at preheating have been reached in \cite{FB}.  The
main point we wish to emphasize is that causality restricts the {\it shape
of the spectrum} of
amplified modes, but not directly the wavelengths
that can be amplified.\footnote{
The convenient approximation sometimes used in
preheating, that the distribution of amplified modes falls as a
spike, $\delta (k - k_{\rm resonance})$,
violates causality, since this requires
that the field fluctuation contain correlations on all length
scales, even for modes with $k/a\gg H$.
}

Preheating in single-field models
is typically restricted to the narrow-resonance regime, and we may expect
much larger
effects in models with multiple scalar fields coupled to the
oscillating
inflaton, because the resonance parameter, $q$, may be much greater than
unity \cite{KLS,BKMT}. The dynamics of such coupled-oscillator systems
are in general chaotic \cite{chaotic} and lead to enhanced particle
production \cite{BT98}.
The multi-field generalization of
equations (\ref{Yeq1})--(\ref{y0i}) can be given as
\cite{kof88,BKMT}
\beqn
&&3H\dot{\Phi}+\left[(k/a)^2+3H^2\right]\Phi = \nonumber\\
&&~~~~~~~-{\ts{1\over2}}\kappa^2\Sigma\left[\dot{\vp}_i(\dvp_i)^{\rd}-
\Phi\dot{\vp}_i^2+V_i\dvp_i\right]\,,  \label{m1}\\
&&(\dvp_i)^{\rd\rd}+3H(\dvp_i)^{\rd}+(k^2/a^2)
\dvp_i = \nonumber\\
&&~~~~~~~ 4\dot{\Phi}\dot{\vp}_i-2V_i\Phi-\Sigma
V_{ij}\dvp_j \,,\label{m2}\\
&&\dot{\Phi}+H\Phi =
{\textstyle{1\over2}}\kappa^2 \Sigma \dot{\varphi}_i
\dvp_i \,,
\label{3}
\eeqn
dropping the mode label $k$, using proper time $t$,
and writing $V_i=\partial V/\partial\vp_i$.
Eq. (\ref{3}) shows that
resonant amplification of field fluctuations is
accompanied by similar behavior of gravitational
fluctuations.
In the single-field case, any resonant growth
occurs with the same
characteristics for metric and field
fluctuations, and stability bands are thus also the same.
In the multi-field case,
this simple relation is broken, and the stability band
structure is much more complicated.
The non-inflaton fields
$\varphi_i$ ($i>1$) grow rapidly under
resonance, while the inflaton $\varphi_1$
is strongly damped. Thus {\em metric fluctuations
$\Phi$ grow more quickly than any of the field
fluctuations $\delta \varphi_i$} due to the quadratic products $\varphi_i
\delta \varphi_i$.
In models which include a substantial broad
resonance regime for the coupled field fluctuations, resonance
parameters are much larger than in the single-field case, typically
in the range  $q\sim 10^2 - 10^6$, rather than $q\leq O(1)$ \cite{exp},
so the associated Floquet indices are much larger \cite{exp,KLS}.

These qualitative remarks are illustrated
in Fig. 1 and amplified in the extensive
simulations in \cite{BKMT},  arising from integrating Eqs.
(\ref{m1})--(\ref{3}) for the
potential $V={1\over2}m^2\vp_1^2+{1\over2}g\vp_1^2\vp_2^2$,
which describes decay of the massive inflaton $\vp_1$
into the boson field $\vp_2$. The graph shows the resonant
amplification and nonlinear growth of gravitational fluctuations
on both super- and sub-hubble scales. Also
clear is the validity of the Floquet index as a characterizer of
growth at strong resonance in an expanding universe (the slopes in the
resonance bands are very nearly
constant). The expansion has the effect of pulling modes through
the resonance bands leading to phases of explosive growth and quiescence.
Note that even with the expansion of the universe included, the Floquet
index is a useful concept at broad resonance  -- the growth of $\Phi_k$ is
almost exactly exponential in the resonance bands, as is evident in Fig.
(1).  A further crucial point, with wide-ranging
implications, is that the non-gravitationally dominated evolution of
perturbations during preheating breaks the scale-invariance of the
primordial  spectrum, as is evident from comparing the evolution of the $k
= 0, 20$ modes in Fig. (1) \cite{BKMT}.

The numerical simulations in \cite{BKMT} reveal that inclusion of the
coupled metric perturbations {\em enhances} the strength of the resonances
in $\delta \varphi_i$, compared with when $\Phi$ is neglected.  This
can be understood by the fact that the gravitational field has negative
specific heat.  For this reason, $\Phi$ does not act as a \lq parasite' or
\lq competitor' with the $\delta \varphi_i$ for the energy of the
oscillating inflaton, but rather can serve as a source or pump for field
fluctuations growth.

Preheating in the narrow resonance and  broad resonance regimes shows
another important qualitative difference: broad-resonance preheating in an
expanding universe proceeds {\it stochastically} \cite{KLS}. The phases of
the amplified field modes are virtually uncorrelated between each
moment when the oscillating inflaton passes through  zero. Both the
field modes and the metric perturbations will have
stochastic driving terms.  It has been shown
that such stochastic terms in general remove all stability bands, so
that {\it all} modes $k\geq0$ grow subject to a parametric resonance,
with, in general, larger characteristic exponents
than in the simple periodic case \cite{noise,BT98}.

Note that once the fluctuations have become strongly nonlinear,
the linearized perturbation equations are no longer valid, and mode-mode
coupling, arising due to convolutions which are ignored in
the linearized Eqs. (\ref{m1})--(\ref{3}), must be included
\cite{KLS}.

Nonlinear gravitational fluctuations
could produce various observable signatures, in particular on the
CMB. Firstly, if the preheating phase is followed by a second
round of inflationary expansion, then sub-hubble
amplified modes would get stretched into
observationally-interesting scales.
Such double-inflation is typical for
most realistic scenarios based on supergravity or  supersymmetry
\cite{sugrasusy}. Secondly, acoustic Doppler peaks at $\ell\geq100$
could also be affected. These are a key
prediction of most inflationary models. In contrast, defect models
often predict no secondary  peaks and a
shift in the position of the first peak,
because the metric perturbations are produced
randomly, receiving out of
phase ``kicks" due
to the mode-mode coupling inherent in the nonlinearity of
defect models \cite{sad}.

In the multi-field, broad-resonance case, stochastic amplification
leads to nonlinear metric fluctuations, producing
mode-mode coupling which drives the
unequal-time correlation function
$\Delta \rightarrow 0$ for $|\tilde\eta-\eta|
> \eta_{\rm c}$, the coherence time of the system.
As $\eta_{\rm c} \rightarrow 0$ we get $\delta$-correlated (white)
noise
\cite{BT98}, memory of initial conditions (and hence coherence) is
lost,
and the field evolution mimics the {\em active,} incoherent evolution
of defects. The crucial question is whether
the nonlinear mode-mode coupling survives local interactions
and persists up to nucleosynthesis and
photon decoupling.
If so, there could be significant limits placed on
inflationary reheating by observed
element abundances and small-scale CMB anisotropies.
By smearing out the Doppler peaks,
surviving nonlinear mode-mode coupling
would greatly reduce the effectiveness
of small-angle CMB observations for differentiating  inflationary
scenarios from defects models.
The implied nonlinearity would lead to hybrid CMB
anisotropies  -- passive and
typically inflationary on large angular scales,
active and defect-like on smaller scales.

Another possible observational consequence
arises from gravitational
bremsstrahlung in scattering of the large  scalar
perturbations \cite{gw}. Combined with our analysis of causality
above, the extra gravitational wave power due to the previously
neglected metric fluctuations
implies that the probability of detection in
instruments such as LIGO may
be  higher than previously estimated, at least for preheating
following chaotic inflation. Nonlinear gravitational fluctuations
could also create the large density contrast necessary to produce
primordial black holes (PBHs) \cite{pbh}, without the need for a
blue spectrum or a large peak in the power spectrum at some $k$.
PBH limits may then be able to constrain the nature of preheating and
the associated amplification of metric perturbations.

Finally,
amplified metric perturbations affect non-thermal
symmetry restoration at preheating \cite{NTSR}, by altering the
inflaton's effective potential via addition of the terms (see
Eq. (10.68) of \cite{MFB}):
$\Delta V = a^2 V_{ \varphi\varphi} \langle \delta \varphi^2 \rangle
+ 2
a^2 V_{ \varphi} \langle \delta \varphi \Phi \rangle$.
The first term was originally considered in \cite{NTSR}, while
the
second is due to the direct (gravitational) coupling between $\dvp$
and $\Phi$. It is of the same perturbative order
as the first term, and is missed if gravitational fluctuations
are neglected in preheating.\\

\noindent{\bf Acknowledgements}

Thanks to Fabrizio Tamburini for very valuable assistance with the
numerics, and to
Juan Garcia-Bellido, Andrew Liddle, Fabrizio Tamburini,
Robert Brandenberger,  Fabio Finelli,  Marco Bruni
and David Wands for discussions.
DK receives  partial support from NSF grant PHY-98-02709.


\newpage
\begin{figure}
\epsfxsize=5.2in
\epsffile{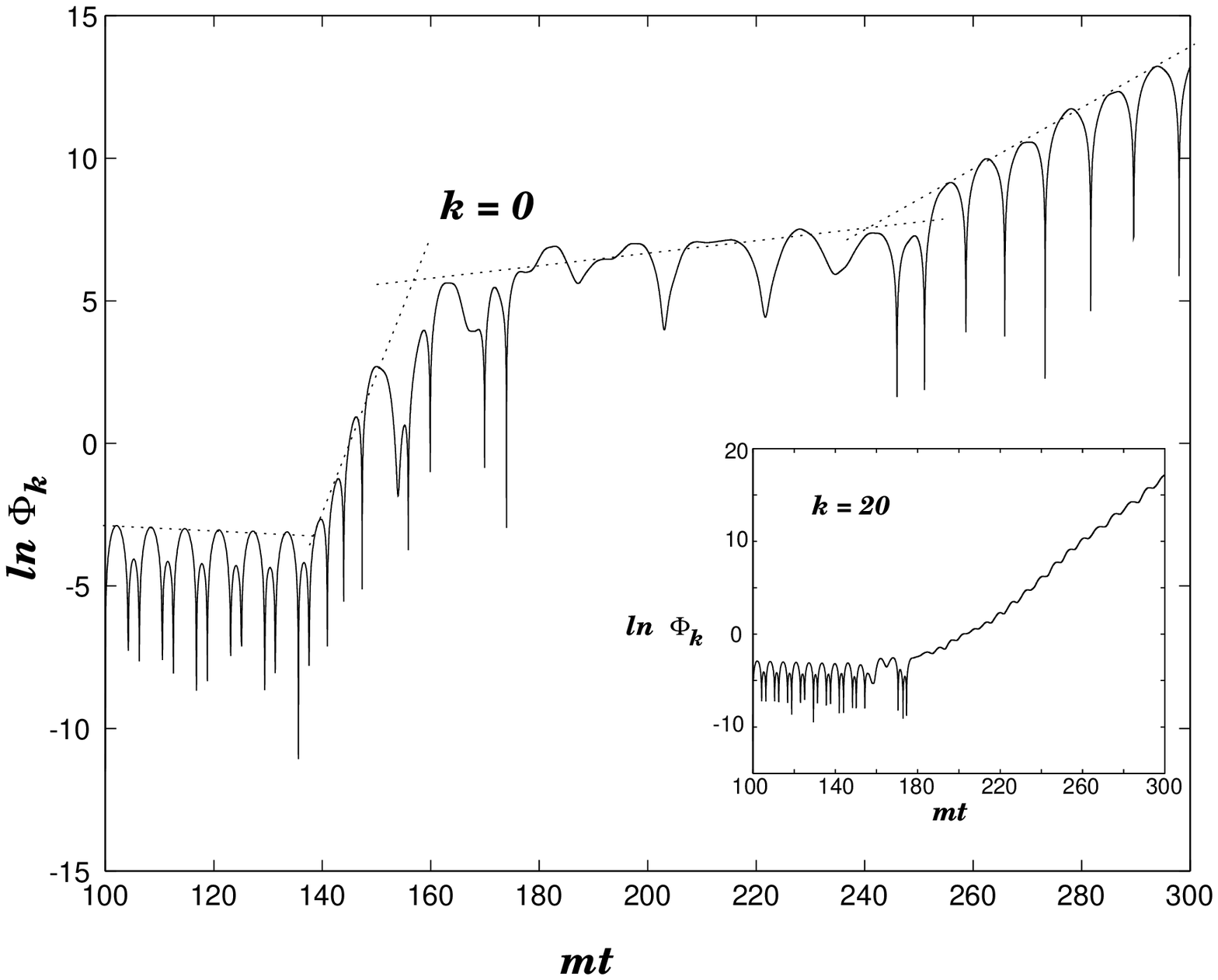}
~\\
\caption{
Metric perturbation evolution in  2-field preheating,
with $q \equiv g
\varphi_{1}^2(t_0)/m^2=8\times10^3$. The main graph shows
the $k=0$ mode, the inset the $k=20$ mode.
Initially, $mt_0= 100$ (well into the small
amplitude phase: $\varphi_{1}(t_0)/mt_0 = 3 \times 10^{-3}$)
and $\Phi_k(t_0)=10^{-5}$, $\dot\Phi_k(t_0)=0$.
{\em The $k=0$  mode becomes nonlinear at $mt \sim 150$ after less
than $10$ inflaton oscillations}  (i.e. well before the end of preheating)
and continues growing without bound. Two strong resonance bands
are evident, with different Floquet indices (given by the slopes of the
dotted lines), for
$140\leq mt \leq 160$ and $250\leq mt \leq 300$.
}
\label{fig:fig1}
\end{figure}


\begin{references}
\bibitem{preh} L. Kofman, A. Linde, and A. Starobinsky, Phys. Rev.
Lett. {\bf 73}, 3195 (1994); Y. Shtanov, J. Traschen, and R.
Brandenberger, Phys. Rev. D {\bf 51}, 5438 (1995);  D. Boyanovsky
et al., Phys. Rev. D {\bf 51}, 4419 (1995) and {\bf 52}, 6805
(1995); M. Yoshimura, Prog. Theo. Phys. {\bf 94}, 8873 (1995); H.
Fujisaki et al., Phys. Rev. D {\bf 53}, 6805 (1996).  Earlier
perturbative studies of resonance included J. Traschen and R. H.
Brandenberger, Phys. Rev. D {\bf 42}, 2491 (1990), and A. Dolgov and D.
Kirilova, Sov. J. Nuc. Phys. {\bf 51}, 172 (1990).
\bibitem{exp} D. Kaiser, Phys. Rev. D {\bf 53}, 1776 (1996); S.Y.
Khlebnikov and I.I. Tkachev, Phys. Rev. Lett. {\bf 77}, 219 (1996)
and {\bf 79}, 1607 (1997), and Phys. Lett. {\bf B390}, 80 (1997); T.
Prokopec and T.G. Roos, Phys. Rev. D {\bf 55}, 3768 (1997); D.
Boyanovsky et al., Phys. Rev. D {\bf 56}, 1939 (1997); I.
Zlatev,
G. Huey, and P.J. Steinhardt, Phys. Rev. D {\bf 57}, 2152 (1998);
B.A. Bassett and S. Liberati, Phys. Rev. D{\bf 58}, 021302 (1998).
\bibitem{KLS} L. Kofman, A. Linde, and A. Starobinsky, Phys. Rev. D
{\bf 56}, 3258 (1997).
\bibitem{metric} H. Kodama and T. Hamazaki, Prog. Theo. Phys. {\bf
96}, 949 and 1123 (1996).
\bibitem{NT} Y. Nambu and A. Taruya,
Prog. Theo. Phys. {\bf 97}, 83 (1997) and Phys. Lett. {\bf B428}, 37
(1998).
\bibitem{BKMT} B.A. Bassett, F. Tamburini, D.I. Kaiser, and R. Maartens,
hep-ph/9901319;
http://www-thphys.physics.
ox.ac.uk/WWW/Users/BruceBassett/reheating/
\bibitem{MFB} V.F. Mukhanov, H.A. Feldman, and R.H.
Brandenberger, Phys. Rep. {\bf 215}, 203 (1992).
\bibitem{gw} S.Y. Khlebnikov and I.I. Tkachev, Phys. Rev. D {\bf
56}, 653 (1997); B.A. Bassett, Phys. Rev. D {\bf 56}, 3429 (1997).
\bibitem{Kandrup} H. E. Kandrup, astro-ph/9903434.
\bibitem{FB} F. Finelli and R. H. Brandenberger, Phys. Rev. Lett. {\bf
82}, 1362 (1999).
\bibitem{PE} M. Parry and R. Easther, Phys. Rev. D {\bf 59}, 061301
(1999).
\bibitem{ABM} V. Mukhanov, L. Abramo, and R. Brandenberger, Phys.
Rev.
Lett. {\bf 78}, 1624 (1997) and Phys. Rev. D {\bf 56}, 3248 (1997).
\bibitem{negg} P.B. Greene, T. Prokopec, and T.G. Roos, Phys. Rev.
D {\bf 58}, 6484 (1997).
\bibitem{inflreviews} A. D. Linde, {\em Particle Physics and Inflationary
Cosmology} (Chur:  Harwood, 1990); E. W. Kolb and M. S. Turner, {\em The
Early Universe} (Redwood City:  Addison-Wesley, 1990);  A. Liddle and D.
Lyth, Phys. Rep. {\bf 231}, 1 (1993).
\bibitem{causal} J. Robinson and B.D. Wandelt, Phys. Rev. D {\bf 53},
618 (1996); N. Turok, Phys. Rev. D {\bf 54}, 3686 (1996) and
Phys. Rev. Lett. {\bf 77}, 4138 (1996).
\bibitem{chaotic} N.J. Cornish and J.J. Levin, Phys. Rev. D {\bf 53},
3022 (1996); R. Easther and K. Maeda, Class. Quantum Grav. {\bf 16}, in press
(1999) (gr-qc/9711035).
\bibitem{BT98} B.A. Bassett and F. Tamburini, Phys. Rev. Lett. {\bf 81},
2630 (1998). 
\bibitem{kof88} L. Kofman and D. Pogosyan, Phys. Lett. {\bf B214},
508 (1988).
\bibitem{noise} V. Zanchin et al., Phys. Rev. D {\bf 57}, 4651
(1998); B.A. Bassett, Phys. Rev. D {\bf 58}, 021303 (1998).
\bibitem{sugrasusy}  J. A. Adams, G.G. Ross, and
S. Sarkar, Nucl. Phys. B {\bf 503} 405 (1997);  R. Jeannerot, Phys.
Rev. D {\bf 53}, 5426 (1996);
D.H. Lyth and E.D.  Stewart, Phys. Rev. Lett. {\bf
75}, 201 (1995).
\bibitem{sad} A. Albrecht et al., Phys. Rev. Lett. {\bf 76}, 1413
(1996); J. Magueijo et al., Phys. Rev. Lett. {\bf 76},  2617
(1996)
and Phys. Rev. D {\bf 54}, 3727 (1996).
\bibitem{pbh}  J. Bullock and J. Primack, Phys. Rev. D {\bf 55}, 7423
(1996).
\bibitem{NTSR} L. Kofman, A. Linde, and A. Starobinsky, Phys. Rev.
Lett.
{\bf 76}, 1011 (1996); I.I. Tkachev, Phys. Lett. {\bf B376}, 35
(1996).

\end{references}
\end{document}